\documentclass[%
 reprint,
 amsmath,amssymb,
 aps,
prl,
floatfix,
]{revtex4-1}

\usepackage{graphicx}
\usepackage{bm}
\usepackage{braket}
\usepackage{hyperref}
\usepackage[usenames]{xcolor}

\DeclareMathOperator{\sgn}{sgn}

\begin{document}

\title{Critical Viscosity of a Fluctuating Superconductor}

\author{Yunxiang Liao}
\author{Victor Galitski}%
\affiliation{Joint Quantum Institute, University of Maryland, College Park, Maryland, 20742, USA}
\affiliation{Condensed Matter Theory Center, University of Maryland, College Park, MD 20742, USA}

\date{\today}

\begin{abstract}
 We consider a  fluctuating superconductor in the vicinity of the transition temperature, $T_c$. The fluctuation shear viscosity is calculated. In two dimensions, the leading correction to viscosity is negative and scales as $\delta \eta(T) \propto \ln(T-T_c)$.   Critical hydrodynamics of the fluctuating superconductor involves two fluids -- a fluid of fluctuating pairs and a quasiparticle fluid of single-electron excitations. The  pair viscosity (Aslamazov-Larkin) term is shown to be zero. The (density of states) correction to viscosity of single-electron excitations  is negative, which is due to fluctuating pairing that results in a  reduction of electron density. Scattering of electrons off of the fluctuations gives rise to  an enhanced quasiparticle scattering and another (Maki-Thomson) negative correction to viscosity.  Our results  suggest that fluctuating superconductors provide a promising platform to investigate low-viscosity electronic media and may potentially host  fermionic/electronic turbulence. Some  experimental probes of two-fluid critical hydrodynamics are proposed such as time-of-flight measurement of turbulent energy cascades in critical cold atom superfluids and  magnetic dynamos in three-dimensional fluctuating superconductors. 
 
\end{abstract}

\maketitle
Motion of classical fluids and astrophysical gases and plasmas is usually described by hydrodynamics. The central  equation of hydrodynamics is the Navier-Stokes equation, which represents a momentum conservation law. In weakly-interacting electronic systems, disorder is the dominant mechanism of momentum relaxation. It strongly breaks  the translational invariance, and the hydrodynamic description, that hinges on the conservation of momentum, is not applicable. In clean and strongly-correlated materials, where the dominant relaxation mechanism is due to interactions, the hydrodynamic description of the electron fluid becomes relevant. This hydrodynamic transport regime has been the subject of much research and interest recently~\cite{transport-1,transport-2,transport-3,transport-4,transport-5,transport-6,transport-7,lucas}. In particular, hydrodynamic electron flows have been reported in experimental studies of graphene~\cite{kumar,graphene2}, Weyl semimetals~\cite{gooth,gooth2017,fu}, and other materials~\cite{flow1,flow2}. 

Viscosity is a central quantity in hydrodynamic theories. It determines the Reynolds number of the flow, which in turn determines its  qualitative type -- laminar or turbulent. The latter turbulent regime is rich with a variety of complicated non-linear phenomena, such as  energy cascades~\cite{kolmogorov1941,obukhov1941}. Turbulence requires large  Reynolds numbers and a low kinematic viscosity. Electron liquids  considered so far all have relatively high viscosity and are far from turbulence regime.  On the theory side, a bound on shear viscosity to entropy ratio has been conjectured~\cite{KSS}, which would limit from below viscosity values possible in electron fluids. 

Here we point out a  class of material --  fluctuating superconductors~\cite{VL}, where it appears possible to  achieve a small shear viscosity and that may be promising candidates for turbulent electronic media. Indeed, a charged superfluid has zero shear viscosity and infinite conductivity. The transition into a superconductor is usually second order and  a critical theory applies in its vicinity, where conductivity~\cite{AL,MT,AHL,VarlamovClean,Altshuler,AslamasovVarlamov,Skvortsov,GL,Dorin,Livanov,Finkelstein,ac_nlsm,Kapitulnik,Kapitulnik2},
thermal conductivity~\cite{varlamovthermal,Huse}, 
Nernst coefficient~\cite{Nernst,Huse,Nerst-F}, 
diamagnetic susceptibility~\cite{VGNPfluct}, 
etc.,
exhibit a singular critical behavior. This paper calculates critical shear viscosity in a  clean, fluctuating two-dimensional superconductor in the vicinity of the superconducting transition temperature, $T_c$. It is shown that the shear viscosity is suppressed by fluctuations.

It is usually not possible to calculate exactly the critical behavior due to fluctuations all the way from high temperatures down to $T_c$ (except in effectively four-dimensional theories, where the parquet/renormalization group technique is asymptotically exact~\cite{VGNPfluct}). However, the Aslamazov-Larkin theory of Gaussian  superconducting fluctuations has a wide regime of formal applicability  and has been shown to be extremely useful in quantitatively explaining experimental data in a variety of fluctuating superconductors. Qualitatively, the Aslamazov-Larkin theory is a two-fluid model involving fluctuating Cooper pairs and electron excitations. The fluctuating Cooper pairs are not condensed and have a finite life-time, but behave much like independent carriers, albeit with a composite structure that is important in correctly evaluating their response to external fields. The Aslamazov-Larkin theory~\cite{VL} involves three key effects in transport: (1)~A negative correction to conductivity due to the reduction of electron density of states (DOS)~\cite{DOS}, which occurs because some electrons are paired. (2)~A positive Aslamazov-Larkin (AL) correction~\cite{AL} due to the  direct conductivity of fluctuating Cooper pairs. Since both their density and life-time diverge at the transition, this correction has a double singularity and usually dominates transport. (3)~The third, usually less singular correction is due to the scattering of electrons off of the fluctuating pairs -- the Maki-Thomson (MT) correction~\cite{MT}. Its sign can be either positive or negative. Fluctuation viscosity can be calculated in a similar way, but the hierarchy of diagrams is different from conductivity, as shown below. In two dimensions, they all have the same type of singular behavior [they all contain the factor $\ln(T-T_c)$, c.f. Ref.~\onlinecite{GL}], but the AL viscosity diagram vanishes in the appropriate limit.

Shear viscosity in a fluid moving with an inhomogeneous velocity is a force per unit area (per unit length in two dimensions) per velocity gradient acting between two fluid elements experiencing the velocity gradient. There are two kinds of terms that contribute to viscosity: scatterings at the boundary between the moving layers that slow down the faster moving ones and drag forces that occur in the presence of long-range interactions. We will consider only short-range interactions and hence the drag viscosity is absent in what follows. The Kubo formula for viscosity $\eta$ has been derived in Refs.~\cite{zubarev,Kadanoff,Hosoya} and reads
\begin{equation}
\label{Kubo}
\begin{aligned}
	K^{R}(\omega)
	=\,&
	-i
	\int_{-\infty}^{+\infty}dt
	\int d^{d}\bm{\mathrm{r}}\,
	e^{i\omega t }
	\Theta(t)
	\braket{
	\left[ 
	\hat{T}_{xy}(\bm{\mathrm{r}},t),
	\hat{T}_{xy}(\bm{\mathrm{0}},0)
	\right] 
	},
    \\
    \eta =\,&
    \lim\limits_{\omega \to 0} \left[ \frac{1}{-i\omega}K^{R}(\omega)\right].
\end{aligned}
\end{equation}
Here, $\hat{T}_{\alpha \beta}$ is the stress-energy tensor operator. 

The stress-tensor is derived  from the continuity relation $\partial_t \hat{j}_\alpha = - \partial_\beta \hat{T}_{\alpha \beta}$, where $\hat{j}_\alpha$ is the $\alpha$'s component of the momentum density operator, $\alpha$ labels spatial axes, and $\partial_\alpha$ is the corresponding derivative. We will consider a clean, interacting electron liquid with the standard Hamiltonian as follows
\begin{equation}
\label{HFL}
\hat{H} =\frac{1}{2m} \int_{\bf r} \partial_\alpha \hat{\Psi}_\sigma^\dagger ({\bf r}) \partial_\alpha \hat{\Psi}_\sigma({\bf r}) + \frac{1}{2} \int_{{\bf r},{\bf r}'} \hat{n}({\bf r}) V({\bf r} -{\bf r}') \hat{n}({\bf r}'),
\end{equation}
where $\hat{\Psi}_\sigma^\dagger ({\bf r})$ and $\hat{\Psi}_\sigma({\bf r})$ are electron field operators creating/destroying electrons with spin $\sigma$ in point ${\bf r}$, and $\hat{n}({\bf r}) =  \hat{\Psi}_\sigma^\dagger ({\bf r}) \hat{\Psi}_\sigma({\bf r})$ is the electron density in point ${\bf r}$. The interaction will be assumed local attraction $V({\bf r}) = - V_0 \delta({\bf r})$, with the appropriate cut-offs used as standard in the BCS theory. From the Heisenberg equations of motion for the current, we find that the local interactions do not contribute to the off-diagonal component of the tensor and its non-interacting form~\cite{Schwinger} can be used
\begin{equation}
\hat{T}_{\alpha \beta}= \! \frac{1}{2m}\left( \partial_\alpha \hat{\Psi}_\sigma^\dagger \partial_\beta \hat{\Psi}_\sigma
\!+\! \partial_\beta \hat{\Psi}_\sigma^\dagger \partial_\alpha \hat{\Psi}_\sigma 
\right).
\end{equation}
Here $\alpha$ and $\beta$ are two arbitrary but different spatial indices.
We will denote the corresponding ``double current'' vertices in the diagrams for viscosity by two  short wavy lines.

	\begin{figure}
	  \vspace{10pt}
	    \centering
	    \includegraphics[width=\linewidth]{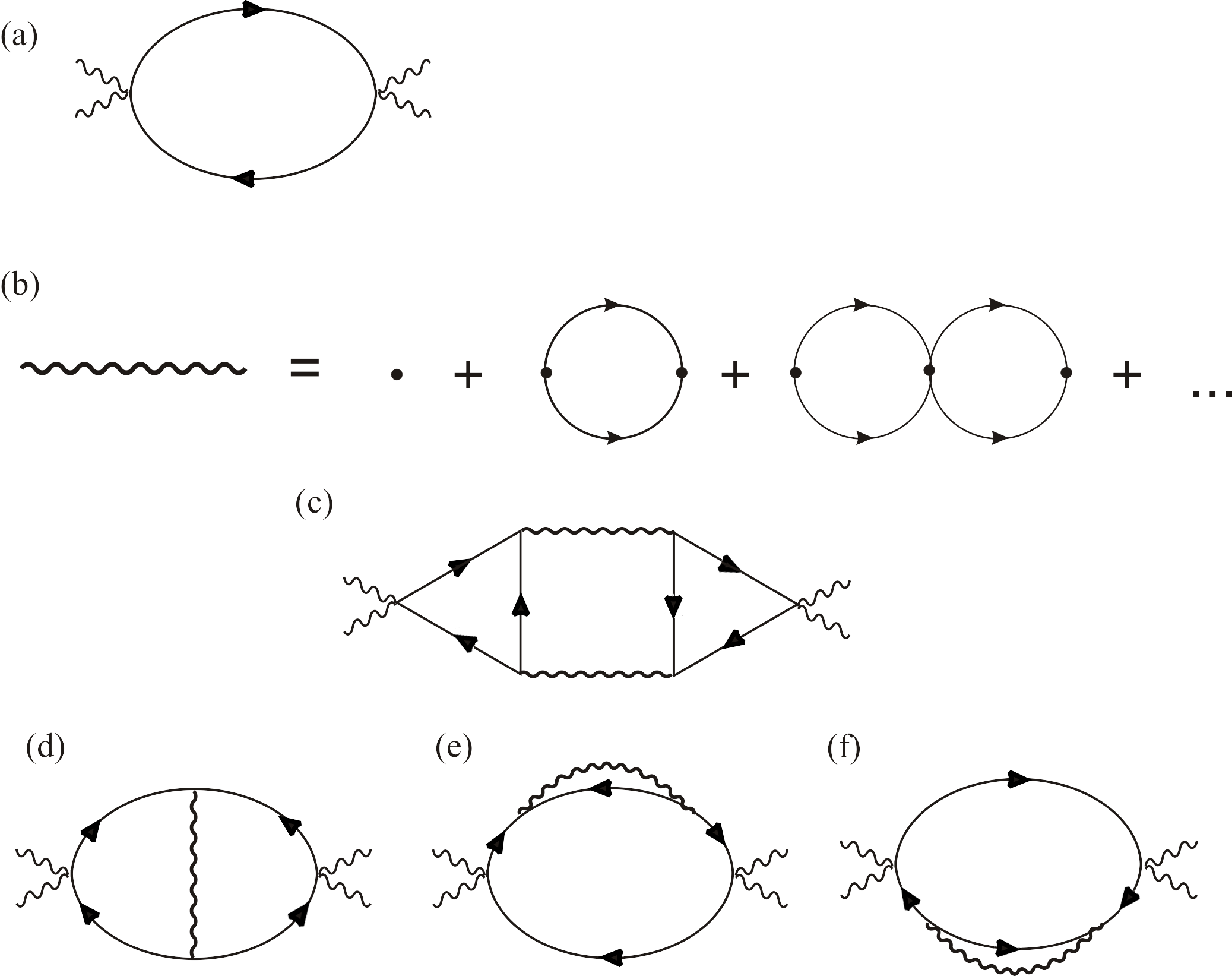}
	    \caption{These diagrams define the main contributions to viscosity discussed and calculated in the main text. (a)~This ``Drude-like diagram'' defines  viscosity of a Fermi liquid with short-range interactions. The interactions give rise to a finite relaxation rate encoded in the Green's functions -- the solid lines. The short double wavy lines correspond to the viscosity vertices $p_x p_y/m$. (b)~The long wavy line is the superconducting fluctuation propagator, see Eq.~\ref{flpropagator}. It diverges at the transition point for $Q=0$.
	    (c)~The Aslamazov-Larkin (AL) diagram for viscosity, which corresponds to viscosity of the fluid of fluctuating pairs. (d)~The Maki-Thomson (MT) diagram for viscosity, which corresponds to scattering of electrons off of the fluctuating pairs. (e) and (f) The density of states (DOS) diagrams for viscosity, which describe deficit of single-electron excitations contributing to $\eta$, because some electrons participate in fluctuating pairing.
	      \label{fig:diagrams}}
	\end{figure}

The viscosity of a Fermi liquid was first discussed qualitatively by Pomeranchuk in 1950~\cite{pomeranchuk}, who argued that it should scale as $\eta\propto T^{-2}$ in three-dimensional metals. This result was later derived more rigorously  by Abrikosov and Khalatnikov~\cite{abrikosov1957,Abrikosov}, who used kinetic equation methods. The simplest way to reproduce this behavior is to consider the ``bubble diagram'' in Fig.~\ref{fig:diagrams}(a) -- the analogue to Drude diagram for viscosity -- with the solid lines representing the Matsubara Green's function, $G^{-1}(\varepsilon_n, {\bf p}) = i \varepsilon_n - \xi_{\bf p} + i\sgn \varepsilon_n /[2 \tau_{\rm FL}(\varepsilon_n)]$ with $\varepsilon_n = (2n+1)\pi T$ being the fermion Matsubara frequencies,  $\xi_{\bf p} = \frac{p^2}{2m} - E_{\rm F}$ is the electron dispersion relative to the Fermi energy, and $\tau_{\rm FL}(\varepsilon_n)$ is the momentum relaxation time. Importantly, here and in what follows, we will assume no disorder and so relaxation is entirely due to interactions. In 3D,  $\tau^{-1}_{\rm 3D FL} \propto T^2$ and in 2D, $\tau^{-1}_{\rm 2D FL} \propto T^2\ln(1/T)$. A  calculation of the na\"{\i}ve ``Drude viscosity diagram'' reproduces the Pomeranchuk-Abrikosov-Khalatnikov scaling
\begin{equation}
\label{visFL}
\eta_{\rm FL}(\omega) \sim  \frac{E_F^2 \nu \tau_{\rm FL}}{1-i \omega \tau_{\rm FL}},
\end{equation}
where $\nu=m/\pi\hbar^2$ is the density of states at the Fermi surface, $m$ is the electron effective mass, and $\omega$ is the external frequency. Note that we have ignored the vertex corrections and also non-local viscosity vertices due to non-local interactions. The expression of   the DC viscosity is proportional to the momentum relaxation time and it reproduces Pomeranchuk's scaling~\cite{pomeranchuk}. At $T=0$, the DC viscosity formally diverges. However, in this limit as well as in a theory with vanishing interactions, the result depends on the order of limits $\tau_{\rm FL} \to \infty$ and $\omega \to 0$, which is dictated by the time-scales in a particular experiment. 

Note that a similar behavior of viscosity occurs in the $\phi^4$ theory, described by the Lagrangian $- {\cal L} = (\partial \phi)^2 + m^2 \phi^2 + \lambda \phi^4$, which was considered by Jeon and Yaffe~\cite{phi4}, who found $\eta_{\phi^4} \propto T^3/\lambda^2$. The viscosity is mass-independent in the leading order and diverges fast as $\lambda \to 0$. However, just like in the Fermi liquid case, for the non-interacting field theory with $\lambda = 0$, a proper order of limits should  be used. 

These results~\cite{phi4} may seem disconcerting, as the perturbative superconducting fluctuation theory is a Gaussian $|\phi|^2$ theory~\cite{VL} (with the complex field $\phi$ playing the role of pair fluctuations, whose ``mass'' is the proximity to the transition), which neglects interactions between the fluctuating pairs (i.e, $\lambda=0$). However, the theory is not Lorentz invariant and the proper fluctuation propagator (see, Eq.~\ref{flpropagator} below) includes  relaxation encoded in its frequency dependence, which is due to the presence of the second fluid -- the single electron excitations. Hence, it is a different effective theory from Ref.~\cite{phi4}. Furthermore, this effective (Ginzburg-Landau) theory has electrons integrated out, which is not appropriate if the flow gradients ``resolve'' the lengthscales smaller than the coherence length (i.e., the pair size). We will assume such ``small-scale'' regime and calculate viscosity using the microscopic Aslamazov-Larkin theory. 

The diagrams for the three processes are presented in Fig.~\ref{fig:diagrams}(c--f). It should be noted that the theory of fluctuations in clean superconductors is highly non-trivial~\cite{Skvortsov}. First, there are several regimes considered in the literature, which depend on the hierarchy  of parameters, $T_c$, $\omega$, the cyclotron frequency in the presence of a field, and the disorder scattering time  $\tau_{\rm imp}$.  The dirty limit $T_c\tau_{\rm imp} \ll 1$ is the simplest, because the Green function blocks in all three diagrams are local, although one has to be careful with including Cooperon modes and treating quantum interference singularities. This limit is irrelevant to our problem. The opposite ultra-clean limit, where the relaxation time is set to infinity [i.e., the Green's functions are taken to be $G^{-1}_0(\varepsilon_n, {\bf p}) = i \varepsilon_n - \xi_{\bf p} $] is the most cumbersome, as it requires regularizations without which it contains pathological results, as discussed in the book of Larkin and Varlamov~\cite{VL}. Namely, the three Green's function blocks in the AL diagram are a non-analytic function of the frequencies in this regime, which presents challenges in using the Matsubara technique.  Ref.~\cite{VarlamovClean} found that there is an exact cancellation of the MT and DOS terms and only the AL diagram survives. This issue was recently critically revisited by Skvortsov et al.~\cite{Skvortsov}, who used the  Keldysh method to circumvent difficulties with analytical continuation. For the sake of completeness, we have presented the calculation of AL viscosity~\cite{Sup} in the ultra-clean limit and encountered similar issues in the Matsubara technique.

However, we argue that this ultra-clean limit is not meaningful and the non-analytic structure of the theory probably reflects inconsistencies in the diagrammatic expansion. Indeed, even in the absence of disorder, interactions (which are necessarily present in a superconductor) give rise to momentum relaxation of Fermi liquid quasiparticles. Hence, there is always a finite relaxation near $T_c$ that must be included in the Green function.  Note that the analogue of the dirty limit does not exist in this  two-fluid fluctuation hydrodynamics, because $\tau_{\rm FL}(T_c) T_c \gg 1$ as long as the Fermi liquid behavior holds. Hence, the Green function blocks are still non-local. However, we find that inclusion of a finite relaxation rate, no matter how small,  straightforwardly regularizes the theory and provides consistent results for all three processes. The details of the calculations are provided in the Supplementary Material~\cite{Sup}. Below we briefly outline the calculation.

The DOS diagram -- see Figs.~\ref{fig:diagrams}(e,f) -- has two key elements:  (i)~First, the fluctuation propagator [the long wavy-line defined in Fig.~\ref{fig:diagrams}(b)],
\begin{equation}
\label{flpropagator}
L({\bf Q},\Omega_k) = - \frac{1}{\nu} \, \frac{1} {\xi^2 Q^2 + \frac{\pi}{8 T_c} |\Omega_k| + (T-T_c)/T_c}
\end{equation}
[with $\xi = \sqrt{\frac{7 \zeta(3)}{32 \pi^2}} v_F/T_c$ and $\zeta(3)\approx1.202$] and (ii)~Second, the four-Green's function block:
\begin{equation}
\label{BDOS}
B_{\rm DOS} (Q,q) = 
T \sum\limits_{\varepsilon_n} \int_{\bf p}\!\! \chi_{\bf p}^2
G_p^2 G_{Q-p} 
\left(
G_{p+q}  
+
G_{p-q} 
\right)
,
\end{equation}
where we introduced for brevity the three-component momenta-frequency: $p=({\bf p},\varepsilon_n)$, $Q=({\bf Q},\Omega_k)$, and $q= ({\bf 0}, \omega_m)$, with the latter representing the external AC frequency ``running'' through the Kubo formula. The function $\chi_{\bf p} = p_x p_y/m$ represents the viscosity vertex (c.f., the current vertex, which is the velocity ${\bf v}= {\bf p}/m$). Here and below, the short-hand notation $\int_{\bf p} \ldots \equiv \int \frac{d^2p}{(2\pi)^2} \ldots $ is used for brevity.

The corresponding Kubo viscosity kernel is
\begin{equation}
\label{KUBODOS}
K_{\rm DOS}(\omega_m) = 2 T \sum\limits_{\Omega_k} \int_{\bf Q} B_{\rm DOS} ({\bf Q}; \Omega_k, \omega_m)  L({\bf Q},\Omega_k).
\end{equation}
The proper analytic continuation $\omega_m \to -i\omega$ and the limit $\omega \to 0$, gives the DC viscosity (Eq.~\ref{Kubo}). The analytical structure of  $B_{\rm DOS} ({\bf Q}; \Omega_k, \omega_m)$ is complicated. However, since we are looking for a singular contribution to viscosity, every power of $\Omega$ and $q$ coming from the block would remove the logarithmic singularity originating from integrating the fluctuation propagator. Hence, we can simply set the bosonic frequency, $\Omega_k$, to zero and focus on the remaining linear-in-$\omega$ term from the block. This greatly simplifies the calculation (note that this simplification is not possible in the absence of regularization). The calculation of the MT contribution -- see, Fig.~\ref{fig:diagrams}(d) -- is similar. One only has to replace
$B_{\rm DOS}$ in Eq.~\ref{KUBODOS} with $B_{\rm MT}$:
\begin{equation}
\label{BMT}
B_{\rm MT} (Q,q) = T \sum\limits_{\varepsilon_n} \int_{\bf p}\!\! \chi_{\bf p} \chi_{{\bf Q} - {\bf p}} G_p G_{p+q} G_{Q-q-p} G_{Q-p}.
\end{equation}

The AL correction, defined in Fig.~\ref{fig:diagrams}(c), is slightly different in that it requires calculation of two triangular blocks
\begin{equation}
\label{BAL}
B_{\rm AL} ({\bf Q}; \Omega_k, \omega_m) = T \sum\limits_{\varepsilon_n} \int_{\bf p} \chi_{\bf p} G_p G_{p+q} G_{Q-p}.
\end{equation}
Note that the angular averaging of the viscosity vertex over the Fermi surface gives zero, unless  we keep contributions proportional to $Q_x Q_y$. This lowers the singularity of the AL diagram down to that of DOS and MT terms (in contrast to the results for conductivity). Furthermore, the AL block is identically zero for $\omega=0$ for any finite $\tau_{\rm FL}$ and hence the product of the two blocks gives rise to the  $\omega^2$ factor. This implies that the DC linear response corresponding to the direct viscosity of fluctuating Cooper pairs vanishes. 

Hence, we are left with the two terms for the viscosity, 
$$
\eta =   \lim\limits_{\omega \to 0} \left\{  \frac{2T}{-i\omega}  \sum\limits_{\Omega_k} \int_{\bf Q} \left[ B_{\rm DOS} ( -i\omega) + B_{\rm MT} ( -i\omega) \right] L(Q)  \right\}.
$$
Putting everything together, we get the main  result of the microscopic calculation -- the fluctuation correction to viscosity,
\begin{equation}
\label{result}
\delta \eta = - \eta_{\rm FL} \frac{ F(T_c \tau_{\rm FL})}{7 \zeta (3)}\, \frac{T_c}{E_F}\, \ln\left(\frac{T_c}{T-T_c}\right),
\end{equation}
where $\eta_{\rm FL}$ is the Fermi liquid viscosity (Eq.~\ref{visFL}), and the $F$-function is 
\begin{equation}
\label{F}
F(\alpha) = 4\pi \alpha \psi'\left(\frac{1}{2}+ \frac{1}{4\pi \alpha} \right) - \frac{1}{2} \psi''\left(\frac{1}{2}+ \frac{1}{4\pi \alpha} \right),
\end{equation}
with $\psi(z)$ being the logarithmic derivative of the $\Gamma$-function. Note that $F(\alpha)$ is strictly positive and hence the leading correction to shear viscosity is strictly negative.

Note that this result (Eq.~\ref{result}), while reliable for a wide range of temperatures can not be trusted all the way down to the transition point.
The leading order perturbation theory is not very informative inside the Ginzburg region [i.e., for $(T-T_c)/T_c \lesssim T_c/E_F = {\rm Gi}$ -- the Ginzburg parameter~\cite{ginzburg,levanyuk}].

	\begin{figure}
	  \vspace{10pt}
	    \centering
	    \includegraphics[width=\linewidth]{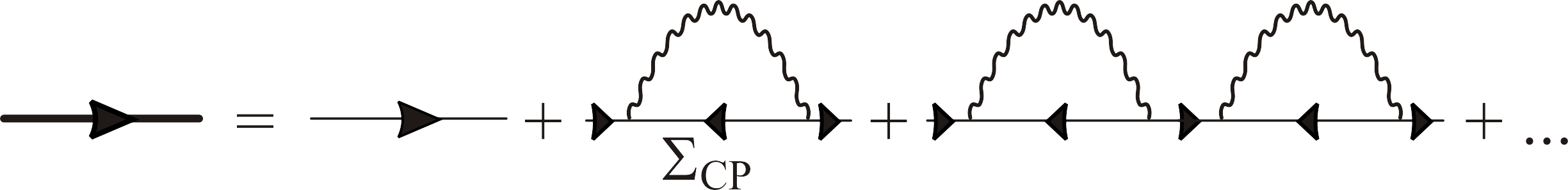}
	    \caption{Single-electron Green's function renormalized by pairing fluctuations. The self-energy (\ref{SE}) gives rise to a strongly enhanced relaxation term calculated in Eq.~(\ref{relax}).
	      \label{fig:selfE}}
	\end{figure}

However, we present phenomenological arguments suggesting that the critical region above the transition is promising to search for electronic turbulence. We do know that at the  transition point the zero-viscosity superfluid forms, which still co-exists with a ``soup'' of Bogoliubov excitations. The phenomenological two-fluid model  below the transition involves a normal fluid that behaves somewhat like an ordinary metal. However,  the two-fluid model right above the transition is markedly different because of strongly enhanced relaxation enabled by the critical uncondensed pairs. The DOS diagram is a precursor to this enhancement. We can resum a subset of diagrams involving the single-electron self-energy due to  pair formation and recombination. 
Consider the self-energy diagram in Fig.~\ref{fig:selfE},
\begin{equation}
\label{SE}
\Sigma_{\rm CP} (p) = T \sum\limits_{\Omega_k} \int _{\bf Q} L({\bf Q}, \Omega_k) G_{Q-p},
\end{equation}
where the fluctuation propagator is given in Eq.~(\ref{flpropagator}). In the leading order it gives
\begin{equation}
\label{relax}
{\rm Im}\, \Sigma_{\rm CP} (p) \equiv -\frac{{\rm sgn} (\varepsilon_n)}{2 \tau_{\rm CP}} 
\sim - {\rm sgn} (\varepsilon_n) \frac{T_c \tau_{\rm FL}}{\nu \xi^2}\, \ln\left(\frac{T_c}{T-T_c}\right).
\end{equation}
Therefore, the dressed Green function has the combined relaxation rate of $\tau^{-1} = \frac{\tau_{\rm FL} + \tau_{\rm CP}}{\tau_{\rm CP}\tau_{\rm FL}}$. If we now calculate the ``Drude-like'' viscosity diagram [see Fig.~\ref{fig:diagrams}(a)] with the dressed Green's function, we obtain the following result for viscosity of the single-electron component above the transition
\begin{equation}
\label{nonpert}
\eta (T \to T_c+) = \frac{\eta_{\rm FL}}{1 + \kappa \ln\left(\frac{T_c}{T-T_c}\right)},
\end{equation}
where $\kappa \sim T_c \tau_{\rm FL}^2/(\nu \xi^2)$. The critical fluctuations above the transition suppress the viscosity of the normal component. We note here that no non-perturbative  theoretical methods exist to reliably describe superconducting fluctuations inside the Ginzburg fluctuation region and Eq.~(\ref{nonpert}) should be viewed as merely an extrapolation of the perturbation theory results. 

Furthermore, the suppression of shear viscosity does not necessarily imply that the hydrodynamic Reynolds number
$$
R \sim  \frac{uL \rho}{\eta}
$$ 
grows, since the latter involves kinematic viscosity given by the ratio of the shear viscosity $\eta$ and the mass density $\rho$ 
(here, $u$ and $L$ are the typical velocity and length-scales of the flow).
We note, however, that the quasiparticle density remains finite even below the transition.
Therefore, vanishing of the shear viscosity  at criticality would indeed imply a small kinematic viscosity and giant hydrodynamic Reynolds number right above the transition. 
The most spectacular consequence of this scenario would be observation of easy-to-create turbulence in the critical region. While the direct measurement of the velocity field and energy spectrum presents a challenge in immediate solid state experiments, it should be quite straightforward in cold fermion superfluids. Indeed, the time-of-flight measurement would provide direct access to the velocity field and enable probe of the Kolmogorov spectrum and potentially inverse energy cascades (see, Ref.~\cite{turbulencereview} for a review). Here, we propose to look for signatures of classical  turbulence in finite temperature neutral fermion superfluids. In particular, low-dimensional such systems  would have a wider critical region and may provide easier access to the regime of interest.

In conclusion, we point out that while our results are specific to two dimensions, three-dimensional fluctuating superconductors may be of special interest from the point of view of exotic (magneto)hydrodynamics as well. In particular, in charged superconductors, the magnetic Reynolds number~\cite{ll}
\begin{equation}
\label{Rm}
R_m = u L \frac{4 \pi \sigma}{c^2}
\end{equation}
is greatly enhanced near the transition for obvious reasons. While the exact critical scaling of the diverging conductivity is unknown in 3D, the Alsamazov-Larkin result provides the following estimate~\cite{VL}
$$
\sigma_{\rm 3D~AL}(\omega, T) = \frac{1}{(1-i\omega \tau)^2} \frac{e^2}{32 \xi} \sqrt{\frac{T_c}{T-T_c}}.
$$
Regardless of critical scaling, the divergence of conductivity at the second-order transition ensures that $R_m$ can be made arbitrarily large. As is known from magnetohydrodynamics and pointed out in our recent Letter~\cite{Turbulence}, this implies instability of differential flows against self-generation of the magnetic field -- the dynamo effect~\cite{Larmor,zeldovich,gilbert1,gilbert2}. Note that while turbulence aids dynamos, it is not necessary. Therefore, critical three-dimensional superconductors above $T_c$ provide a promising playground to attempt observation of self-exciting dynamos in the solid state laboratory. The existing dynamo experiments~\cite{e-dynamos,e1-dynamos,e2-dynamo,e3-dynamo} involve fast rotating classical conducting fluids in large containers (needed to increase the $uL$ factor in  Eq.~\ref{Rm}). However, $R_m$ can be made large in critical superconductors regardless of $uL$.  The simplest experiment, which would mimic experimental classical hydrodynamic dynamos, would therefore involve fast rotation of the sample in the close proximity to superconducting $T_c$ and looking for signatures of the dynamo instability -- a spontaneously generated magnetic field.

\begin{acknowledgments}
		We are grateful to Axel  Brandenburg,  Dam Thanh Son, Sergey Syzranov,  and Andrey Varlamov for useful discussions. 
This research was supported by  US-ARO (contract No. W911NF1310172) (Y.L.), DOE-BES (DESC0001911) (V.G.), and the Simons Foundation.
\end{acknowledgments}
	
\bibliography{main}

\end{document}


\title{Critical Viscosity of a Fluctuating Superconductor\\
	Supplemental Material}

\author{Yunxiang Liao}
\author{Victor Galitski}%
\affiliation{Joint Quantum Institute and Condensed Matter Theory Center, University of Maryland, College Park, Maryland 20742, USA}

	
	\maketitle

In this supplemental material, we show the derivation of the fluctuation corrections to viscosity in the vicinity of the superconducting transition temperature $T_c$, which consists of the Aslamazov-Larkin (AL), the density of states (DOS) and the Maki-Thompson (MT) corrections. 
The viscosity is given by the retarded correlation function $K^{R}(\ww)$ of the stress-energy tensor $T_{xy}$ [see Kubo formula Eq.~(1)]. 
Here, we use the Matsubara technique and calculate the Matsubara correlation function of the stress-energy tensor:
\begin{align}\label{Seq:G}
\begin{aligned}
	&K(\wm)
	=\,
	-
	\int_{0}^{\beta}d \tau
	\int d^{d}\rb
	e^{i\wm \tau }
	\braket{
		T_{xy}(\rb,\tau)
		T_{xy}(\vex{0},0)
	}
	=\,
	-
	\frac{1}{\beta^3}
	\suml{\en,\en'}\,
	\intl{\pb,\pb'}\,
	\chi_{\pb} \chi_{\pb'}
	\braket{	
		\bPsi_{\sigma} (p)
		\Psi_{\sigma} (p+q)
		\bPsi_{\sigma'} (p'+q)
		\Psi_{\sigma'} (p')
	},
\end{aligned}
\end{align}
from which the retarded correlation function can be obtained by analytical continuation $K^{R}(\ww)=K(\wm \rightarrow -i\ww)$. 
Here, as in the main text, 
we use the notation:
$p=({\bf p},\varepsilon_n)$, $q= ({\bf 0}, \omega_m)$, $Q=({\bf Q},\Omega_k)$,  and $\int_{\bf p} (\ldots) \equiv \int \frac{d^2p}{(2\pi)^2} (\ldots) $.
The viscosity vertex is defined as $\chi_{\bf p} \equiv p_x p_y/m$.

\section{I.\ The Aslamazov-Larkin Correction}\label{Sec:AL1}

In this section, we calculate the Aslamazov-Larkin (AL) correction to viscosity. The diagram in Fig. 1(c) gives the following contribution to the Matsubara correlation function $K(\wm)$ of the stress-energy tensor:
\begin{align}\label{Seq:AL}
	\begin{aligned}
	K_{\rm AL}(\wm)
	=\,&
	-
	\frac{4}{\beta}
	\suml{\Omega_k}
	\intl{\qb}\,
	L(\qb,\ok)
	L(\qb,\ok+\wm)
	B_{\rm AL}^2(\qb,\ok,\wm),
\end{aligned}
\end{align}
where $B_{\rm AL}$ gives the amplitude of the three Green's functions block 
\begin{align}\label{Seq:BAL}
\begin{aligned}
	B_{\rm AL}(\qb,\ok,\wm)
	=\,&
	\frac{1}{\beta}\suml{\en}
	\intl{\pb}
	\chi_{\pb}
	\G (p)\G (p+q) \G (Q-p),
\end{aligned}
\end{align}
and $L$ is the fluctuation propagator [Eq.~(5)] represented diagrammatically by the long wavy line [see Fig.~1(b)].
We note that $\G$ is the one-particle Green's function with Fermi liquid relaxation rate $\tau_{\rm FL}^{-1}$:
\begin{align}
\begin{aligned}
	&\G(\pb,\en)
	=
	\dfrac{1}{i\ten-\xi_{\pb}},
\end{aligned}
\end{align}
where $\xi_{\pb} \equiv p^2/2m-E_{\rm F}$ and $\ten$ is defined as
\begin{align}
	\ten \equiv \en + \frac{1}{2\tau_{\rm FL}} \sen.
\end{align}
Note here we ignore the energy dependence of $\tau_{\rm FL}$ and treat it as a finite and temperature dependent constant in the following calculation.

To calculate the value of three Green's functions block $B_{\rm AL}$, we first expand it in terms of momentum $\qb$,
\begin{align}
\begin{aligned}
	B_{\rm AL}(\qb,\ok,\wm)
	=\,&
	\frac{1}{\beta}\suml{\en}
	\intl{\pb}
	\chi_{\pb}
	\G (\pb,\en)\G (\pb,\en+\wm) 
	\\
	\times &
	\left[ 
	\G (\pb,\ok-\en)
	+
	\left(-\frac{\pb \cdot \qb}{m}  + \frac{Q^2}{2m}\right)\G^2 (\pb,\ok-\en)
	+
	\left(\frac{ \pb \cdot \qb}{m}\right)^2
	\G^3 (\pb,\ok-\en)
	\right].
\end{aligned}
\end{align}
Only the third term in the above expression is nonvanishing after angular integration over the Fermi surface, and we have
\begin{align}\label{Seq:BAL1}
\begin{aligned}
	B_{\rm AL}(\qb,\ok,\wm)
	=\,&
	\nu
	\braket{
		\chi_{\pb}
		\left(\frac{\pb \cdot \qb}{m} \right)^2
	}_{\msf{FS}}
	\frac{1}{\beta}
	\suml{\en}
	\int_{-\infty}^{\infty}
	d \xi
	\frac{1}{i\te_{k-n}-\xi}
	\frac{1}{i\te_{m+k-n}-\xi}
	\frac{1}{\left( i\te_{n}-\xi\right)^3},
	\end{aligned}
\end{align}
where we have made the transformation $\en \rightarrow \ok-\en$.
$\nu=m/\pi \hbar^2$ is the density of states at the Fermi surface.
The bracket $\braket{...}_{\msf{FS}}$ denotes the averaging over the Fermi surface, and $\braket{
	\chi_{\pb}
	\left( \frac{\pb \cdot \qb}{m} \right)^2
	}_{\msf{FS}}$ is given by
\begin{align}
\begin{aligned}
	&\braket{
			\chi_{\pb}
			\left(\frac{\pb \cdot \qb}{m} \right)^2
		}_{\msf{FS}}
	=\,
	\frac{1}{4} m \vf^4 Q_x Q_y.
\end{aligned}
\end{align}
Using the Cauchy's residue theorem, we find, for nonzero $\wm \neq 0$, $\xi-$integration in Eq.~\ref{Seq:BAL1} leads to
\begin{align}
\begin{aligned}
	B_{\rm AL}(\qb,\ok,\wm)
	=\,&	
	\frac{1}{4} \nu m \vf^4 Q_x Q_y
	\frac{1}{\beta}
	\suml{\en}
	\left\lbrace 
	\frac{ 2\pi i \sgn \en \Theta (\en \e_{n-k} )}
	{\left[  \wm+ \frac{\tau_{\rm FL}^{-1}}{2} (\se_{n} -\se_{n-k-m}) \right]  (2\en-\ok+ \tau_{\rm FL}^{-1} \sen )^3} 
	\right.
	\\
	&\left.
	-\frac{ 2\pi i \sgn \en \Theta (\en \e_{n-m-k} )}
	{\left[ \wm+ \frac{\tau_{\rm FL}^{-1}}{2} (\se_{n-k} -\se_{n}) \right] (2\en-\ok-\wm+\tau_{\rm FL}^{-1} \sen)^3}
	\right\rbrace,
\end{aligned}
\end{align}
where $\Theta$ denotes the Heaviside step function.
The summation over fermionic frequencies $\en=(2n+1)\pi/\beta$ in the above equation can be evaluated in terms of $\psi-$function which is the logarithmic derivative of the $\Gamma-$function. For $\wm>0$ and $\Omega_k \geq 0$, we have
\begin{align}\label{Seq:BAL2}
\begin{aligned}
	B_{\rm AL}(\qb,\ok,\wm)
	=\,&
	-i
	\nu
	m \vf^4 Q_x Q_y
	 \frac{\beta^2}{ 2^8 \pi^2 } 
	 \\
	 \times &
	\left\lbrace 
	\frac{1}{\wm}
	\left[
	\psi^{(2)}(\frac{1}{2}+\frac{\ok}{4\pi T}+\frac{\wm}{2\pi T}+\frac{\tau^{-1}_{\rm FL}}{4\pi T})
	+
	\psi^{(2)}(\frac{1}{2}+\frac{\ok}{4\pi T}+\frac{\tau^{-1}_{\rm FL}}{4\pi T})
	-
	2
	\psi^{(2)}(\frac{1}{2}+\frac{\ok}{4\pi T}+\frac{\wm}{4 \pi T}+\frac{\tau^{-1}_{\rm FL}}{4\pi T})
	\right]
	\right.
	\\
	&+
	\left.
	\frac{  1 }{\left( \wm + \tau^{-1}_{\rm FL}  \right)}
	\left[ 
	\psi^{(2)}(\frac{1}{2}+\frac{\ok}{4\pi T}+\frac{\tau^{-1}_{\rm FL}}{4\pi T})
	-
	\psi^{(2)}(\frac{1}{2}+\frac{\ok}{4\pi T}+\frac{\wm}{2\pi T}+\frac{\tau^{-1}_{\rm FL}}{4\pi T})
	\right]  
	\right\rbrace .
\end{aligned}
\end{align}

In the vicinity of critical temperature $T_c$, due to the singular structure of the fluctuation propagator $L$ , the leading contribution to $K(\wm)$ comes from the fluctuation propagator $L$ rather than from the block $B_{\rm AL}$. For this reason, we approximate $B_{\rm AL}$ by its value at $\Omega_k=0$.
Analytical continuation $\wm \rightarrow -i\ww$ and expanding in terms of small $\ww$ leads to
\begin{align}\label{Seq:BAL4}
\begin{aligned}
	&B_{\rm AL}(\qb,\ok=0,-i \ww)
	=\,
	\nu
	m \vf^4 Q_x Q_y
	 \frac{\beta^3}{ 2^9 \pi^3 } 
	\tau_{\rm FL}
	\left[ 
	\psi^{(3)}(\frac{1}{2}+\frac{\tau_{\rm FL}^{-1}}{4 \pi T})
	-
	\frac{\tau_{\rm FL}^{-1}}{8 \pi T}\psi^{(4)}(\frac{1}{2}+\frac{\tau_{\rm FL}^{-1}}{4 \pi T})
	\right] \ww
	+ O(\ww^2),
\end{aligned}
\end{align}
which vanishes in the $\ww \rightarrow 0$ limit.
Inserting the above expression into Eq.~(\ref{Seq:AL}) and Eq.~(1), one can draw the conclusion that the AL correction to viscosity vanishes.

We also note that the value of the block $B_{\rm AL}$ directly calculated at $\wm=0$ vanishes, consistent with the result given by Eq.~(\ref{Seq:BAL4}).
At $\wm=0$, the $\xi-$integration in Eq.~\ref{Seq:BAL1} becomes
\begin{align}\label{Seq:BAL5}
\begin{aligned}
&\int_{-\infty}^{\infty}
d \xi
\frac{1}{(i\te_{k-n}-\xi)^2}
\frac{1}{\left( i\te_{n}-\xi\right)^3}
=\,
\frac{-6 \pi i  \sgn \en \Theta (\en \e_{n-k} )}
{(2\en-\ok+ \tau_{\rm FL}^{-1} \sen )^{4}}.
\end{aligned}
\end{align}
Summing the above expression over all fermionic frequencies $\en$ leads to $0$,
consistent with the result obtained by analytically continuing $\wm \rightarrow -i\ww$ and taking $\ww \rightarrow 0$ limit.

\section{II.\ The Aslamazov-Larkin Correction in the $\tau_{\rm FL} \to \infty$ Limit}

For the sake of completeness, we present here calculations in the limit where momentum relexation is absent from the outset. I.e., $\tau_{\rm FL}^{-1}=0$. As mentioned in the text,
we believe that this limit is unphysical, as in any superconductor, interactions guarantee presence of momentum relaxation at any finite temperature and these relaxation terms
provide an important regularization to the Aslamazov-Larkin diagram. The calculations below are presented to simplify comparison with the previous literature \cite{VarlamovClean} and to illustrate issues with the analytical continuation that inevitably arise. 

In the  $\tau_{\rm FL} \to \infty$ limit, for positive $\wm>0$ and zero $\ok=0$, the block $B_{\rm AL}$ is given by
\begin{align}\label{Seq:AL_B_C}
\begin{aligned}
	&B_{\rm AL}(\qb,\ok=0,\wm)
	=\,
	-i \nu m \vf^4  Q_x Q_y 	
	\frac{\beta^2}{2^7 \pi^2 } 
	\frac{1}{\wm} 
	\left[ 
	\psi^{(2)} (\frac{1}{2})
	-
	\psi^{(2)} (\frac{1}{2}+\frac{\wm}{4\pi T})
	\right],
\end{aligned}
\end{align}
which can be obtained from Eq.~(\ref{Seq:BAL2}) by setting $\tau_{\rm FL}^{-1}$ and $\ok$ to $0$.
After analytically continuing $\wm \rightarrow -i\ww$ and taking the DC ($\ww \rightarrow 0$) limit, we find $B_{\rm AL}$ is nonvanishing and an imaginary quantity:
\begin{align}\label{Seq:BAL-1}
\begin{aligned}
	B_{\rm AL}(\qb,\ok=0,-i\ww)
	=\,
	i \nu m \vf^4  Q_x Q_y 	
	\frac{\beta^3}{ 2^9 \pi^3 } 
	\psi^{(3)} (\frac{1}{2})
	+O(\ww).
\end{aligned}
\end{align}
By contrast, setting $\tau_{\rm FL}^{-1}=0$ in Eq.~(\ref{Seq:BAL5}), we find $B_{\rm AL}$ vanishes at $\wm=0$, as in the finite $\tau_{\rm FL}^{-1}$ case.
This shows that $B_{\rm AL}(\wm)$ breaks the  analyticity at $\wm=0$ in the  $\tau_{\rm FL} \to \infty$ limit, which can be resolved by employing the Keldysh technique~\cite{VarlamovClean,Skvortsov}.

\begin{figure}
	\vspace{10pt}
	\centering
	\includegraphics[width=0.5\linewidth]{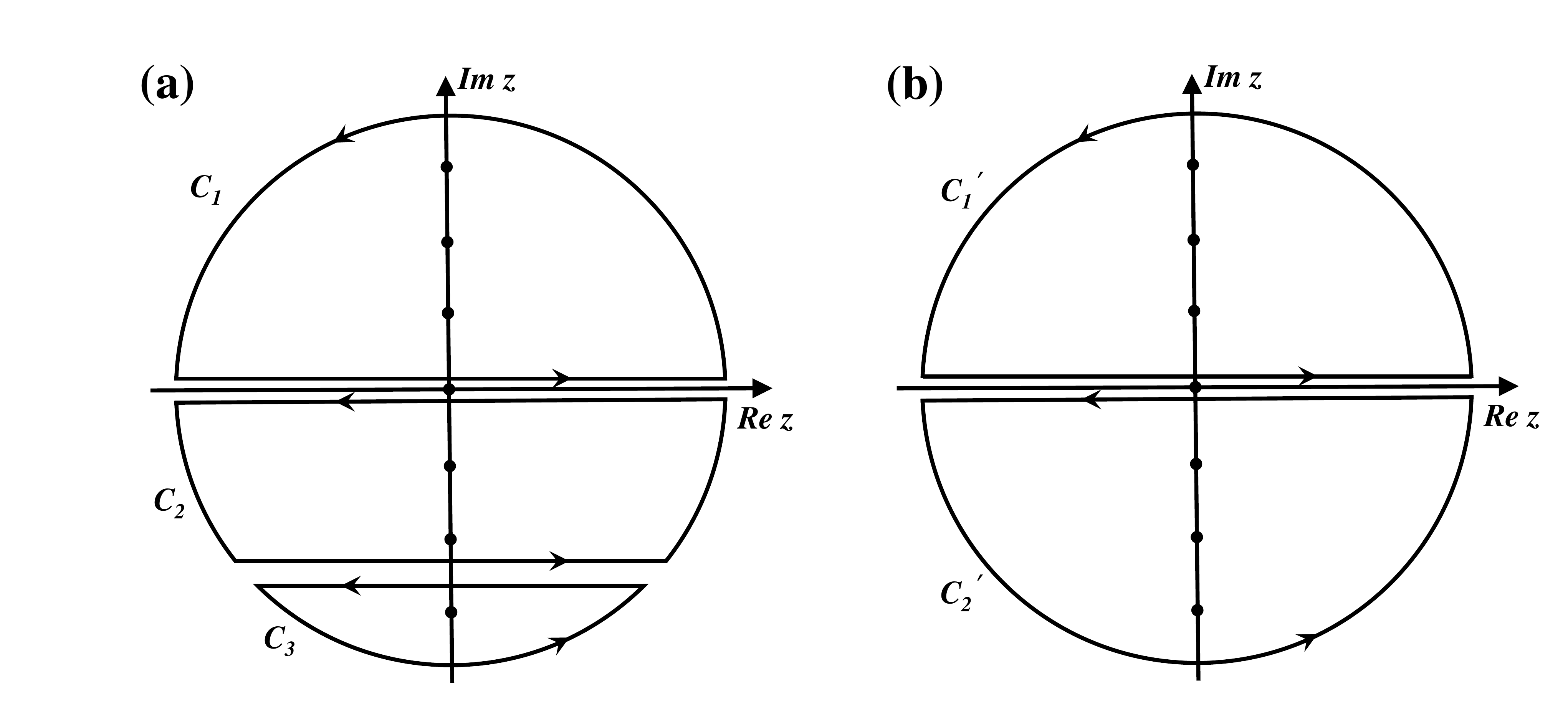}
	\caption{Fig. (a) [(b)] depicts the integration contour $C$ [$C'$] used in the  in Eq.~(\ref{Seq:KAL}) [Eq.~(\ref{Seq:KDOS})] to calculate the sum over bosonic frequencies $\ok$.
		\label{fig:contours}}
\end{figure}

Inserting the Eq.~(\ref{Seq:BAL-1}) into Eq.~(\ref{Seq:AL}) leads to a negative AL correction to viscosity. The proof is as follows.
The summation over bosonic frequencies $\ok$ in Eq.~(\ref{Seq:AL}) can be rewritten as a contour integral~\cite{VL}
\begin{align}\label{Seq:KAL}
\begin{aligned}
	&K_{\rm AL} (\wm)
	=\,
	-
	4
	\intl{\qb}\,
	\tilde{B}_{\rm AL}^2 (\qb)	
	\oint_{C} \frac{dz}{4 \pi i}
	\coth (\frac{\beta z}{2})	
	L(\qb,-iz)
	L(\qb,-iz+\wm),
\end{aligned}
\end{align}
where we approximate $B_{\rm AL}(\qb,\ok,-i\ww)$ with Eq.~(\ref{Seq:BAL-1}) which has been represented here by $\tilde{B}_{AL} (\qb)$. 
The contour $C$, depicted in Fig.~\ref{fig:contours}(a), is composed of three closed contours $C_1$, $C_2$ and $C_3$.
It avoids the cuts at $\Ima z=0$ and $\Ima z=-\wm$ where the analyticities of the functions $L(\qb,-iz)$ and $L(\qb,-iz+\wm)$ are broken, respectively.
The contour integral can be further reduced to
\begin{align}
\begin{aligned}
	K_{\rm AL}(\wm)
	=\,&
	-
	4
	\intl{\qb}\,
	\tilde{B}_{AL}^2 (\qb)	
	\left\lbrace 
	\int_{-\infty}^{\infty} \frac{dz}{4 \pi i}
	\coth (\frac{\beta z}{2})	
	\left[ L^{R}(\qb,-iz)-L^{A}(\qb,-iz) \right] 
	L^{R}(\qb,-iz+\wm)
	\right.
	\\
	&\left.
	+
	\int_{-\infty-i\wm}^{\infty-i\wm} \frac{dz}{4 \pi i}
	\coth (\frac{\beta z}{2})	
	L^{A}(\qb,-iz) 
	\left[ L^{R}(\qb,-iz+\wm)-L^{A}(\qb,-iz+\wm)\right] 
	\right\rbrace 	
	\\
	=\,&
	-
	4
	\intl{\qb}\,
	\tilde{B}_{AL}^2 (\qb)	
	\int_{-\infty}^{\infty} \frac{dz}{4 \pi i}
	\coth (\frac{\beta z}{2})	
	\left[ L^{R}(\qb,-iz)-L^{A}(\qb,-iz) \right] 
	\left[ 	L^{R}(\qb,-iz+\wm)+L^{A}(\qb,-iz-\wm) \right].
\end{aligned}
\end{align}
Here in the second equality, we have applied the shift $z \rightarrow z-i\wm$ in the second integral and also used the fact that $\coth (\beta (z+i\wm)/2)=\coth (\beta z/2)$.

Analytical continuation $\wm \rightarrow -i \ww$ leads to
\begin{align}\label{Seq:KAL-1}
\begin{aligned}
	K_{\rm AL}^{R} (\omega)
	=\,&
	-i \frac{2}{\pi} \ww
	\intl{\qb}\,
	\tilde{B}_{\rm AL}^2 (\qb)	
	\int_{-\infty}^{\infty} dz
	\coth (\frac{\beta z}{2})	
	\,\partial_z \left[ \operatorname{Im} L^{R}(\qb,-iz) \right]^2
	\\
	=\,&
	-i \frac{\beta}{\pi}  \ww
	\intl{\qb}\,
	\tilde{B}_{\rm AL}^2 (\qb)	
	\int_{-\infty}^{\infty} dz
	\dfrac{1}{\sinh^2 (\frac{\beta z}{2})}
	\left[  \Ima L^{R}(\qb,-iz)	\right]^2,
\end{aligned}
\end{align}
where only the linear term in $\ww$ is kept in $K_{\rm AL}^{R}(\ww)$. The higher order terms vanish after being inserted into the Kubo formula [Eq.~(1)] evaluated at $\ww \rightarrow 0$ limit, while the $\ww^0$ term is canceled by the corresponding contributions from other diagrams.
In the second equality, an integration by parts has been used.

Inserting the expressions for $\tilde{B}_{\rm AL} (\qb)$ [Eq.~\ref{Seq:BAL-1}] together with that of $L(\qb,-iz)$ [Eq.~(5)] into Eq.~(\ref{Seq:KAL-1}), and using the Kubo formula Eq.~(1), we find the AL correction to viscosity in the  $\tau_{\rm FL} \to \infty$ limit is given by
\begin{align}
\begin{aligned}
\delta \eta_{\rm AL}
=\,&
\frac{ \beta}{\pi}
\intl{\qb}\,
\tilde{B}_{\rm AL}^2 (\qb)	
\int_{-\infty}^{\infty} dz
\dfrac{1}{\sinh^2 (\frac{\beta z}{2})}
\left[  \Ima L^{R}(\qb,-iz)	\right]^2
=\,	
\dfrac{\pi}{4\nu^2}
\intl{\qb}\,
\tilde{B}_{\rm AL}^2 (\qb)
\dfrac{1}{\left[ \xi^2 Q^2 +(T-T_c)/T_c \right] ^3}
\\
=\,&
-\dfrac{\pi^2}{2^{25}} m^2 \vf^8 \beta^6
\frac{1}{\xi^6}
\ln \left( \dfrac{\xi^2 \Lambda_q^2}{(T-T_c)/T_c } \right)
=\,
-\dfrac{\pi^8}{2^{10}7^3\zeta(3)^3} m^2 \vf^2
\ln \left( \dfrac{T_c}{T-T_c}\right).
\end{aligned}
\end{align}
Here in the 2nd equality, since the most singular contribution comes from the region $z \ll T$, we have approximated $\sinh (\beta z/2)$ by $\beta z/2$ 
and carried out the integration in $z$. $\Lambda_q \sim \xi^{-1}$ represents the ultraviolet cutoff of the momentum integration, where the coherence length $\xi$ is given by $\sqrt{\frac{7\zeta(3)}{32\pi^2}}\vf/T$. 

It is curious that in the relaxation-less case, the direct viscosity of fluctuating Cooper pairs is negative. The sign comes from the fermionic vertices, which contain information about the composite structure of the pairs. This corresponds to a flow, where the lengthscales of the viscosity gradient are smaller than the Cooper pair size (the negative sign may reflect the tendency towards the formation of pairs with counter propagating electrons). The composite nature of the pairs makes the calculation markedly different from the featureless Ginzburg-Landau critical theory or $\phi^4$ theory, which would describe viscosity of fundamental bosons. Furthermore, as emphasized in the main text and in the beginning of the section, the results of the calculation for viscosity depend on the order of limits $\tau_{\rm FL} \to\infty$ and $\omega\to 0$. In the case of the Aslamazov-Larkin diagram, taking the $\tau_{\rm FL} \to\infty$ limit first (as considered in this section) is unphysical and the results of the previous section [Sec.~I] should be used instead (i.e., $\delta \eta_{\rm AL} = 0$). 

\section{III.\ The Density of States Correction}

The contributions from diagrams in Figs.~1 (e) and (f) are given by $K_{\rm DOS{+}}(\wm)$ and $K_{\rm DOS{-}}(\wm)$, respectively,
\begin{align}\label{Seq:DOS}
\begin{aligned}
	K_{\rm DOS{\pm}}(\wm)
	=\,&
	\frac{2}{\beta}
	\suml{\ok}
	\intl{\qb}\,
	L(\qb,\ok)
	B_{\rm DOS{\pm}}(\qb;\ok,\wm),
	\\
	B_{\rm DOS{\pm}}(\qb;\ok,\wm)
	=\,&
	\frac{1}{\beta}\suml{\en}
	\intl{\pb}
	\chi_{\pb}^2
	\G^2 (p) \G (p \pm q) \G(Q-p).
\end{aligned}
\end{align}
Their sum gives the density of state (DOS) contribution to Matsubara correlation function $K(\wm)$.
In the following, we will calculate the viscosity correction from the diagram depicted Figs.~1 (f). It is straightforward to verify that the diagram in Figs.~1 (e) gives an identical correction to viscosity.

In the vicinity of critical temperature $T_c$, the most singular contribution to $K_{\rm DOS}(\wm)$ comes from the fluctuation propagator $L$. As a result, one can ignore the $Q$ dependence of $B_{\rm DOS+}(Q,\wm)$ and set $Q$ to $0$,
\begin{align}\label{Seq:BDOS1}
\begin{aligned}
	B_{\rm DOS+}(Q=0,\wm)
	=\,
	\nu 
	\braket{\chi_{\pb}^2}_{\msf{FS}}
	\frac{1}{\beta}\suml{\en}
	\int_{-\infty}^{\infty}
	d \xi
	\frac{1}{(i\ten-\xi)^2}
	\frac{1}{i\te_{n + m} -\xi}
	\frac{1}{i\te_{-n}-\xi},
\end{aligned}
\end{align}
Here averaging $\chi_{\pb}^2$ over the Fermi surface gives
\begin{align}
\begin{aligned}
	&\braket{\chi_{\pb}^2}_{\msf{FS}}
	=\,
	\frac{1}{8} m^2 \vf^4.
\end{aligned}
\end{align}
and the $\xi-$integration leads to
 \begin{align}
\begin{aligned}
	&
	\int_{-\infty}^{\infty}
	d \xi
	\frac{1}{(i\ten-\xi)^2}
	\frac{1}{i\te_{n + m} -\xi}
	\frac{1}{i\te_{-n}-\xi}
	\\
	=\,
	&
	\frac
	{ 2 \pi \sgn \en \Theta \left[ -\en \e_{n+m} \right] }
	{( \wm-\tau_{\rm FL}^{-1}\sgn \en)^2(2 \en + \wm )}
	-
	\frac
	{  2 \pi  \sgn \en }
	{(2\en +\tau_{\rm FL}^{-1}\sgn \en)^2[2\en + \wm +\frac{\tau_{\rm FL}^{-1}}{2}(\sgn \en+\sgn \e_{n+m}) ]}	,
\end{aligned}
\end{align}
at nonzero $\wm$. For positive $\wm$, summation over fermionic frequencies $\en$ leads to
\begin{align}
\begin{aligned}
	&B_{\rm DOS{+}}(Q=0,\wm)
	=\,	
	 -\nu m^2 \vf^4
	\frac{1}{2^4}
	\left\lbrace 
	\frac{1}{\wm^2}	
	\left[ 
	\psi(\frac{1}{2}+\frac{\tau_{\rm FL}^{-1}}{4 \pi T})
	+\psi(\frac{1}{2}+\frac{\wm}{2 \pi T}+\frac{\tau_{\rm FL}^{-1}}{4 \pi T})
	-2\psi(\frac{1}{2}+\frac{\wm}{4 \pi T}+\frac{\tau_{\rm FL}^{-1}}{4 \pi T})
	\right] 
	\right.
	\\
	&\,\,
	\left.
	+
	\frac{1}{(\wm+\tau_{\rm FL}^{-1})^2}	
	\left[ 
	\psi(\frac{1}{2}+\frac{\tau_{\rm FL}^{-1}}{4 \pi T})
	-\psi(\frac{1}{2}+\frac{\wm}{2 \pi T}+\frac{\tau_{\rm FL}^{-1}}{4 \pi T})
	\right] 
	+
	\frac{\dfrac{\tau_{\rm FL}^{-1}}{2\pi T}}{2\wm(\wm+\tau_{\rm FL}^{-1})}
	\left[ 
	\psi^{(1)}(\frac{1}{2}+\frac{\tau_{\rm FL}^{-1}}{4 \pi T})
	-\psi^{(1)}(\frac{1}{2}+\frac{\wm}{2 \pi T}+\frac{\tau_{\rm FL}^{-1}}{4 \pi T})
	\right]	
	\right\rbrace.
\end{aligned}
\end{align}
We then analytical continue $\wm \rightarrow -i \ww$, and expand in terms of $\ww$ :
\begin{align}\label{Seq:BDOS2}
\begin{aligned}
	B_{\rm DOS{+}}(Q=0,-i\ww)
	=\,&
	\frac{\beta^2}{2^6\pi^2} 
	\nu m^2 \vf^4
	\left\lbrace 
	\frac{1}{4} \psi^{(2)}(\frac{1}{2}+\frac{\tau_{\rm FL}^{-1}}{4 \pi T})
	\right.
	\\
	&
	\left.
	-i\ww
	\tau_{\rm FL}
	\left[ 
	\frac{2 \pi T}{\tau_{\rm FL}^{-1}}
	\psi^{(1)}(\frac{1}{2}+\frac{\tau_{\rm FL}^{-1}}{4 \pi T})
	-
	\frac{1}{2}
	\psi^{(2)}(\frac{1}{2}+\frac{\tau_{\rm FL}^{-1}}{4 \pi T})
	+
	\frac{\tau_{\rm FL}^{-1}}{16 \pi T}
	\psi^{(3)}(\frac{1}{2}+\frac{\tau_{\rm FL}^{-1}}{4 \pi T})
	\right] 
	\right\rbrace.
\end{aligned}
\end{align}
Later, we will show later the $\ww^0$ term here cancels with the same type of contributions from the other DOS diagram  [$B_{\rm DOS-}$, see Fig.~1(c)] and the MT diagram [Fig.~1(d)], and therefore this term will be ignored from now on.

We note that the $\ww$ dependence of $K^{R}_{\rm DOS+}(\ww)$ only comes from $B_{\rm DOS+}(\wm \rightarrow -i\ww)$ which is approximated by its value at $Q=0$.
Similar to the AL calculation, we reexpress the summation over $\ok$ as an contour integral [see Fig.~\ref{fig:contours}(b) for an illustration of the integration contour $C'$]:
\begin{align}\label{Seq:KDOS}
\begin{aligned}
	K^{R}_{\rm DOS+}(\ww)
	=\,&		
	2B_{\rm DOS+}(-i\ww)
	\intl{\qb}\,
	\oint_{C'} \, \frac{dz}{4 \pi i} \coth (\frac{\beta z}{2})
	L(\qb,-iz)
	=\,
	2 B_{\rm DOS+}(-i\ww)
	\intl{\qb}
	\int_{-\infty}^{\infty} \, \frac{dz}{2 \pi } \coth (\frac{\beta z}{2})
	\text{Im} L^{(R)}(\qb,-iz).
\end{aligned}
\end{align}
Using Eq.~(1) and the fact that the other DOS diagram gives an identical contribution to viscosity, we find a negative DOS correction to viscosity
\begin{align}\label{Seq:DOS1}
\begin{aligned}
	\delta \eta_{\rm DOS}
	=\,&
	-\frac{\beta}{ 2^6 \pi^3  }
	  m^2 \vf^4 \tau_{\rm FL}
	  \frac{1}{\xi^2}
	F_{\rm DOS}(T \tau_{\rm FL})
	\ln \left[ \frac{\xi^2 \Lambda_q^2 }{(T-T_c)/T_c}\right]
	\\
	=\,&
	-	\frac{1}{  14\zeta(3)\pi  }
	T  m^2 \vf^2 \tau_{\rm FL}
	F_{\rm DOS}(T \tau_{\rm FL})
	\ln \left[ \frac{T_c}{T-T_c}\right],
\end{aligned}
\end{align}
where $F_{\rm DOS}(x)$ is defined as
\begin{align}\label{Seq:DOS2}
\begin{aligned}
	F_{\rm DOS}(x)
	\equiv \,&
	2\pi x
	\psi^{(1)}(\frac{1}{2}+\frac{1}{4\pi x})
	-
	\frac{1}{2}
	\psi^{(2)}(\frac{1}{2}+\frac{1}{4\pi x})
	+
	\frac{1}{16 \pi x}
	\psi^{(3)}(\frac{1}{2}+\frac{1}{4\pi x})
	,
\end{aligned}
\end{align}
positive for any $x > 0$.

\section{IV.\ The Maki-Thompson Correction}

The calculation of the Maki-Thompson (MT) correction is performed in a similar manner as was done for the DOS correction.
The corresponding diagram, illustrated in Fig.~1(d), gives the following contribution to the Matsubara correlation function $K(\wm)$:
 
\begin{align}\label{Seq:MT}
\begin{aligned}
	K_{\rm MT}(\wm)
	=\,&
	\frac{2}{\beta}
	\suml{\ok}
	\intl{\qb}\,
	L(\qb,\ok)
	B_{\rm MT}(\qb,\ok,\wm),
	\\
	B_{\rm MT}(\qb,\ok,\wm)
	=\,&
	\frac{1}{\beta}\suml{n}
	\intl{\pb}
	\chi_{\pb} \chi_{\qb-\pb}
	\G (p) \G (p+q) 
	\G(Q-p)  \G(Q-p-q).
\end{aligned}
\end{align}

Similar to the DOS calculation, it is sufficient to approximate $B_{\rm MT}$ by its value at $Q=0$:
\begin{align}
\begin{aligned}
	&B_{\rm MT}(Q=0,\wm)
	=\,
	\nu 
	\braket{\chi_{\pb} \chi_{-\pb}}_{\msf{FS}}
	\frac{1}{\beta}\suml{\en}
	\int_{-\infty}^{\infty}
	d\xi
	\frac{1}{i\ten-\xi}
	\frac{1}{i\te_{n+m} -\xi}
	\frac{1}{i\te_{-n}-\xi}
	\frac{1}{i\te_{-m-n}-\xi}
	\\
	=\,&
	 \frac{1}{8} \nu m^2 \vf^4
	 \frac{1}{\beta}\suml{\en}
	\left\lbrace 
	\frac
	{2 \pi \sgn \en \Theta \left[ -\en \e_{n+m} \right] }
	{(\wm-\tau_{\rm FL}^{-1} \sgn \en )
		(2 \en + \wm )
		(2 \en + 2\wm -\tau_{\rm FL}^{-1}\sgn \en) }
	\right.
	\\
	&
	\left.
	+
	\frac
	{2 \pi  \sgn \en  }
	{(2\en +\tau_{\rm FL}^{-1}\sgn \en)
		\left[ 2\en + \wm + \frac{\tau_{\rm FL}^{-1}}{2} (\sgn \en +\sgn \e_{n+m})\right] 
		 \left[ \wm+\frac{\tau_{\rm FL}^{-1}}{2} (-\sgn \en +\sgn \e_{n+m})\right] }	
	\right.
	\\
	&
	\left.
	-
	\frac
	{2 \pi  \sgn \en \Theta \left[ \en \e_{n+m}\right] }
	{(2\en + \wm +\tau_{\rm FL}^{-1} \sgn \en)
		( 2\en +2\wm + \tau_{\rm FL}^{-1} \sgn \en )  \wm}	
	\right\rbrace,
\end{aligned}
\end{align}	
where in the second equality, we have used the fact $\braket{\chi_{\pb} \chi_{-\pb}}_{\msf{FS}}$ equals $\braket{\chi_{\pb}^2}_{\msf{FS}}$ and carried out the $\xi-$integration using the Cauchy's residue theorem.

Summing over fermionic frequencies $\en$ and expressing the result in terms of the $\psi-$ function, we have
\begin{align}
\begin{aligned}
	B_{\rm MT}(Q=0,\wm)
	=\,
	 \frac{1}{8} \nu m^2 \vf^4
	&
	\left\lbrace 
	\frac{1}{(\wm+\tau^{-1}_{\rm FL})^2}
	\left[ 
	-\psi(\frac{1}{2}+\frac{\tau_{\rm FL}^{-1}}{4 \pi T})
	+\psi(\frac{1}{2}+\frac{\wm}{2 \pi T}+\frac{\tau_{\rm FL}^{-1}}{4 \pi T})
	\right] 
	\right.
	\\
	&\,+\left.
		\frac{1}{\wm^2}
		\left[
		-\psi(\frac{1}{2}+\frac{\tau_{\rm FL}^{-1}}{4 \pi T})
		-\psi(\frac{1}{2}+\frac{\wm}{2 \pi T}+\frac{\tau_{\rm FL}^{-1}}{4 \pi T})
		+2\psi(\frac{1}{2}+\frac{\wm}{4 \pi T}+\frac{\tau_{\rm FL}^{-1}}{4 \pi T} )
		\right]	
	\right\rbrace,
\end{aligned}
\end{align}
for positive $\wm$.
After analytically continuing $\wm \rightarrow -i \ww$ and expanding in terms of $\ww$, we have
\begin{align}\label{Seq:BMT}
\begin{aligned}
	&B_{\rm MT}(Q=0,-i\ww)
	=\,
	\frac{\beta^2}{2^5\pi^2}
	\nu m^2 \vf^4
	\left\lbrace 
	-\frac{1}{4}\psi^{(2)}(\frac{1}{2}+\frac{\tau_{\rm FL}^{-1}}{4 \pi T})
	-i\ww
	\tau_{\rm FL}
	\left[ 
	\frac{2 \pi T}{\tau_{\rm FL}^{-1}}
	\psi^{(1)}(\frac{1}{2}+\frac{\tau_{\rm FL}^{-1}}{4 \pi T})
	-
	\frac{\tau_{\rm FL}^{-1}}{16 \pi T}
	\psi^{(3)}(\frac{1}{2}+\frac{\tau_{\rm FL}^{-1}}{4 \pi T})
	\right] 
	\right\rbrace.
\end{aligned}
\end{align}
We note that the $\ww^0$ term in $B_{\rm MT}$ cancel with the corresponding contribution from the DOS diagrams ($B_{\rm DOS+}+B_{\rm DOS-}$).

The MT and DOS contributions only differ through the corresponding block value $B(-i\ww)$ [see Eqs.~(\ref{Seq:DOS}) and~(\ref{Seq:MT})]. As a result, one can deduce the value of MT correction to viscosity using Eqs.~(\ref{Seq:BMT}),~(\ref{Seq:DOS1}) and~(\ref{Seq:DOS2}):
	\begin{align}
\begin{aligned}
	\delta \eta_{\rm MT}
	=\,
	-	\frac{1}{  14\zeta(3)\pi  }
	T  m^2 \vf^2
	\tau_{\rm FL}
	F_{\rm MT}(T \tau_{\rm FL} )
	\ln \left[ \frac{T_c}{T-T_c}\right],	
	\\
	F_{\rm MT}(x)
	\equiv \,
	2\pi x
	\psi^{(1)}(\frac{1}{2}+\frac{1}{4 \pi x})
	-
	\frac{1}{16 \pi x}
	\psi^{(3)}(\frac{1}{2}+\frac{1}{4 \pi x})
	.
\end{aligned}
\end{align}
Due to the fact that $F_{\rm MT}(x) > 0$ for any $x > 0$, we find the MT diagram also gives a negative correction to viscosity, similar to that of DOS.

Combining the contributions from AL, DOS and MT, we find the total first-order fluctuation correction to viscosity:
\begin{align}
\begin{aligned}
	\delta \eta
	=\,&
	-
	\frac{1}{  14\zeta(3)\pi  }
	T  m^2 \vf^2
	\tau_{\rm FL}
	F(T \tau_{\rm FL})
	\ln \left[ \frac{T_c}{T-T_c}\right],
	\\
	F(x)
	=\,&
	F_{\rm DOS} (x)+ F_{\rm MT} (x) 
	=\,
	4\pi x
	\psi^{(1)}(\frac{1}{2}+\frac{1}{4 \pi x})
	-
	\frac{1}{2}
	\psi^{(2)}(\frac{1}{2}+\frac{1}{4 \pi x}),
\end{aligned}
\end{align}
which can be rewritten as Eqs.~(10) and~(11).

\bibliography{main}